# INTERCYCLE AND INTRACYCLE VARIATION OF HALO CME RATE OBTAINED FROM SOHO/LASCO OBSERVATIONS

**Fithanegest Kassa Dagnew[1,2,3], Nat Gopalswamy[2], Solomon Belay Tessema[1], Sachiko Akiyama[2,3], Seiji Yashiro[2,3], and Tesfay Yemane Tesfu[1]**

[1] Ethiopian Space Science and Technology Institute (ESSTI), Entoto Observatory and Research Center (EORC), Addis Ababa, Ethiopia

[2] NASA Goddard Space Flight Center, Greenbelt, MD, USA

[3] The Catholic University of America, Washington DC, USA

ABSTRACT

We report on the properties of halo coronal mass ejections (HCMEs) in solar cycles 23 and 24. We compare the HCMEs properties between the corresponding phases (rise, maximum, and declining) in cycles 23 and 24 in addition to comparing those between the whole cycles. Despite the significant decline in the sunspot number (SSN) in cycle 24, which dropped by 46% with respect to cycle 23, the abundance of HCMEs is similar in the two cycles. The HCME rate per SSN is 44% higher in cycle 24. In the maximum phase, cycle-24 rate normalized to SSN increased by 127% while the SSN dropped by 43%. The source longitudes of cycle-24 HCMEs are more uniformly distributed than those in cycle 23. We found that the average sky-plane speed in cycle 23 is ~16% higher than that in cycle 24. The size distributions of the associated flares between the two cycles and the corresponding phases are similar. The average speed at a central meridian distance (CMD) $\geq 60^0$ for cycle 23 is ~28% higher than that of cycle 24. We discuss the unusual bump in HCME activity in the declining phase of cycle 23 as due to exceptional active regions that produced many CMEs during October 2003 to October 2005. The differing HCME properties in the two cycles can be attributed to the anomalous expansion of cycle-24 CMEs. Considering the HCMEs in the rise, maximum and declining phases, we find that the maximum phase shows the highest contrast between the two cycles.

# 1. INTRODUCTION

Howard et al. (1982) reported the observation of a halo coronal mass ejection (HCME) for the first time. They described it as a halo of excess brightness completely surrounding the occulting disk and propagating radially outward. Properties of HCMEs were derived from the observations made by the Solwind instrument on board the P78-1 satellite during 1979-1985 (Howard et al. 1985). A significant number of HCMEs were later recorded by the Large Angle and Spectrometric Coronagraph's (LASCO) C2 and C3 telescopes (Brueckneer et al. 1995) on board the Solar and Heliospheric Observatory (SOHO) mission (Gopalswamy et al. 2003; Zhao and Webb 2003; St.Cyr 2005). HCMEs are stated as full and asymmetric (F- and A-type) halos (Gopalswamy et al. 2003; Gopalswamy et al. 2007; Gopalswamy et al. 2010a). Basesd on their source longitudes, HCMEs are classified as disk (F-type) and limb (A-type) halos. Disk halos are HCMEs whose sources are with in $45^0$ central meridian distance (CMD) whereas limb halos are those whose source longitudes are with a CMD between $45^0$ and $90^0$ (Gopalswamy 2009; Gopalswamy et al. 2010b). HCMEs account for ~3% of all CMEs. Only about 10% of halos originate close to the limb (Gopalswamy et al. 2015a). Front-sided HCMEs are a major cause of severe geomagnetic storms and have important space weather implications (St.Cyr et al. 2000; Zhao and Webb 2003; Michalek et al. 2006; Gopalswamy et al. 2007).

Gopalswamy et al. (2014) reported that CMEs of cycle 24 expand anomalously compared to those in cycle 23. They compared limb CMEs (CMD >60 deg) between the two cycles and found that for a given CME speed, the cycle-24 CMEs are significantly wider. They also found a higher fraction of halos among the limb CMEs in cycle 24 than in cycle 23. They suggested that the anomalous expansion is caused by the reduced total pressure in the heliosphere. Petrie (2015) reported a statistically significant increase in the rate of CME detections with angular width > $30^0$ per sunspot number (SSN) for cycle 24 compared to cycle 23. The relative number of CMEs between the two cycles was found to be dependent on the widths, but fast and wide CMEs showed a definite decline in cycle 24 (Gopalswamy et al. 2015b). By conducting a statistical study of a significant data set from two CME catalogs, Compagnino et al. (2017) found that the number of CMEs observed during the maximum of cycle 24 was higher or comparable to the one during cycle 23. Michalek et al. (2019) conducted a statistical analysis of CME detection rates during cycles 23 and 24 and suggested that the higher CME rate in cycle 24 is caused by a significant decrease of total heliospheric pressure and the changed magnetic pattern of solar corona. Lamy et al. (2019) performed a statistical analysis of CMEs covering nearly two complete solar cycles (cycle 23 and 24) and stated that CME rates were relatively larger during cycle 24 than cycle 23.

By comparing the properties of space weather events between cycle 24 and cycle 23, a substantial increase in the fraction of halos was found among CMEs associated with a solar energetic particle (SEP) events (Gopalswamy 2012) and interplanetary type II radio bursts (Gopalswamy et al. 2019). Gopalswamy et al. (2015a) have compared HCMEs in cycle 24 (2008 December–2014 December) with those in cycle 23 (1996 May–2002 June) corresponding to the first 73 months in each cycle. Regardless of the significant drop of SSN, the HCME abundance relative to SSN was higher in cycle 24. They found that the distribution of source locations (longitudes and latitudes) in cycle 24 significantly differ from those in cycle 23. The longitude distribution of halos were much flatter in cycle 24 than in cycle 23. They reported that the kinematics and flare size distributions between the two cycles are similar where the difference in the average speeds (frontside HCMEs) between the two cycles is less than 10%. The peculiar behavior of HCMEs was attributed to the weak state of the heliosphere in cycle 24.

In this paper, we compare the properties of HCMEs in solar cycles 24 (2008 Dec 1–2019 Mar 31) and 23 (1996 May 1–2006 Aug 31) over an equal-length period of 124 months. Since we now have almost the whole of cycle 24, we also compare the HCME properties in the corresponding phases (rise, maximum, and declining) of the two cycles.

The previous study by Gopalswamy et al. (2015a) was done by considering only the rising and maximum phases of the solar cycles. We were motivated to find out what the differences would be if we include the declining phases and perform whole cycle comparisons. We perform intercycle and intracycle comparsions: the rise, maximum, and declining phases of the two cycles.

Automatic CME detection algorithms can partially or completely miss CME events and even record false detections(Hess and Colaninno(2017)). Moreover,the results of automatic catalogs like the SEEDS (Solar Eruptive Event Detection System, **http://spaceweather.gmu.edu/seeds/**), CACTUS (Computer-Aided CME Tracking, **http://sidc.oma.be/cactus/**), CORIMP(Coronal image processing, **http://alshamess.ifa.hawaii.edu/CORIMP/**), and ARTEMIS (Automatic Recognition of Transient Events and Marseille Inventory from Synoptic maps, **http://cesam.lam.fr/lascomission/ARTEMIS/**) are inconsistent with each other and with the manual catalog (Richardson et al. 2015; Webb et al. 2017, Lamy et al. 2019). Hess and Colaninno (2017) reported a pronounced divergence among the three catalogs (SEEDS, CACTUS, and CORIMP) in the period between 2001 and 2004. The automatic detection algorithms miss the majority of HCMES and the catalogs have recorded too few of them (Gopalswamy et al. 2020). Lamy et al. 2019 have shown strong divergence in the HCME numbers and speeds between the four catalogs (SEEDS,CACTUS,ARTEMIS,and CORIMP),for example, CACTUS reports HCME numbers and speeds much larger than SEEDS and ARTEMIS while SEEDS reported very few halos and even none during cycle 23. Gopalswamy et al.(2010a) reported that ARTEMIS and SEEDS didn't detect any full halos while CACTUS detected 12.6 % of all full halos.

Therefore, it is important to complete this whole-cycle HCME comparison that will serve as a bench mark to compare HCMEs from automatic detection algorithms.

## 2. OBSERVATIONS

We used data (https://cdaw.gsfc.nasa.gov/CME_list/halo/halo.html; Gopalswamy et al. 2010a) from the catalog that compiles all HCMEs manually identified from SOHO/LASCO images within the C2 and C3 field of view (FOV). The HCME catalog contains all the essential parameters needed for our study. The HCME appearance time is the first appearance of the HCME in the LASCO/C2 FOV. HCMEs are identified by their appearance in C3 FOV because some CMEs become full halos only in the larger FOV of C3. The sky-plane speed of a CME is the apparent speed measured by tracking the fastest moving section of the CME in coronagraph images. The images are two dimensional projections of the white light emission on the plane of the sky. The sky-plane speeds are obtained using a linear fit to the height time measurements made at the measurement position angle (MPA) where a CME seems to move the fastest (Gopalswamy et al. 2010a). The measured speeds are not the real speeds as sources of HCMEs are usually closer to the disk centers. The sky-plane speeds need to be corrected for projection effects to get the actual speeds. The space speed is the actual speed of the CME in 3D space which is essential to understand the CME kinematics in the heliosphere. It is obtained from a cone model using the sky-plane speed and an assumed CME half width (Xie et al. 2004, Gopalswamy 2009). The sky-plane speeds of HCMEs depend on the longitude of the solar source (Gopalswamy et al. 2007; Gopalswamy et al. 2010a). The HCME source locations are the heliographic coordinates of the associated eruption region. Source locations of the associated H-α flare are obtained from the Solar Geophysical Data (SGD) listing. When such data are not available, the source information is obtained from inner coronal images, microwave images from the Nobeyama radioheliograph or H-alpha images (Gopalswamy et al. 2007; Gopalswamy et al. 2010a). We used JavaScript movies that combine LASCO images superposed on EUV images with the GOES X-ray light curves to confirm the solar source and the associated flares. We considered frontsided HCMEs in the comparison of the source longitude and latitude distributions between the two cycles and the corresponding phases. We filled the gaps in the catalog (space speeds) using the relation $V_{Rad} = 1.1*V_{LASCO} +156$ ($V_{Rad}$ = space speed and $V_{LASCO}$ = sky-plane speed of HCMEs) derived by Gopalswamy et al. (2015a).

We make use of the SSN data from the on line Sunspot Index and Long-term Solar Observations (SILSO, http://www.sidc.be/silso). We referred the Carrington rotation (CR) start and stop times (http://umtof.umd.edu/pm/crn/CARRTIME.HTML) for the number of HCMEs per CR.

## 3. ANALYSIS AND RESULTS

In this section, we analayze the HCME abundance, rates per month and per SSN, source locations, speeds, and flare size distributions. We compare these properties between the two cycles and the corresponding phases. Solar cycle 24 has come to its end and cycle 25 spots have started appearing. Therefore, we have full two solar cycles to compare, especially the phases.

### 3.1. Halo CME Rate

Figure 1 shows the number of HCMEs as a function of time binned over a CR period. There were 387 HCMEs in cycle 23 and 318 in cycle 24 observed during the first 124 months in each cycle. If we consider the interruption time in SOHO operations with the assumption of the same HCME rate of occurrence, the number of HCMEs in cycle 23 will be 403 but then the HCME rates per month and per SSN remain the same. Although Gopalswamy et al. (2015a) also considered corrections on the halo count due to smaller LASCO data gaps, we take such effects are negligible. Disregarding the four months without data in cycle 23, the HCME rate per month in cycle 23 is 3.23 and 2.56 in cycle 24. There were only 9 remaining HCMEs in cycle 23 outside the study period with a monthly rate of 0.33 indicating that comparing the first 124 months between the cycles is the same as comparing the whole cycles.

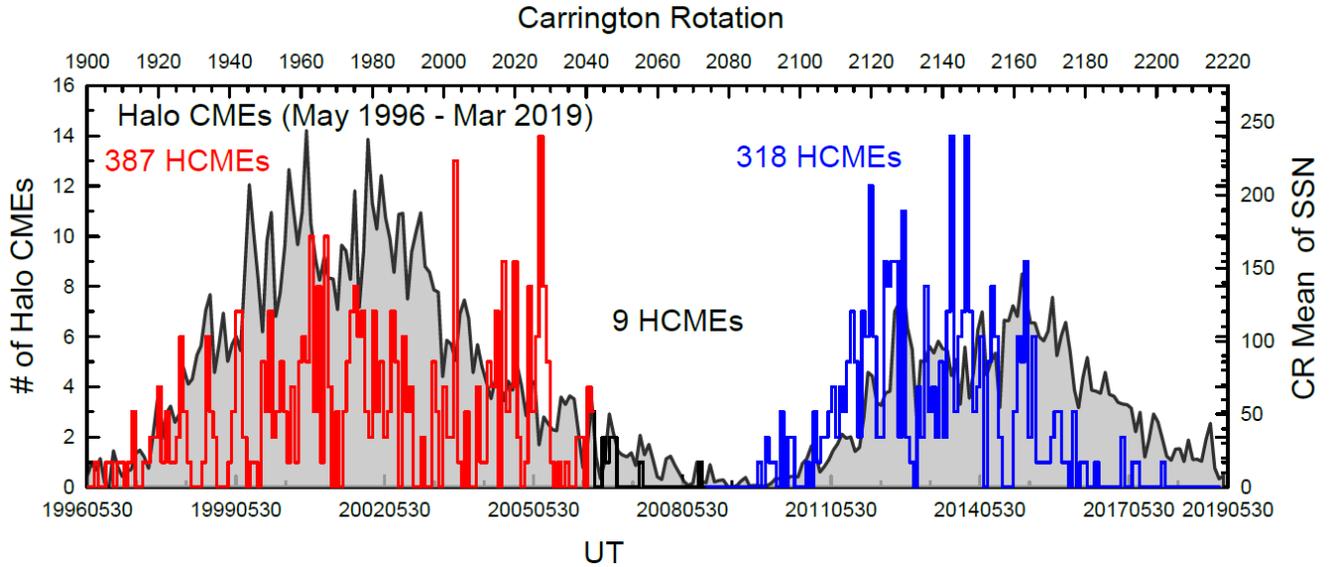

**Figure 1:** The number of halo CMES (HCMEs) as a function of time binned over a Carrington rotation (CR) periods. The red (cycle 23) and blue (cycle 24) histograms correspond to the first 124 months in each cycle. The black histogram in between red and blue correspond to 9 HCMEs observed in cycle 23 after the first 124 month until the end of cycle 23. The grey color refers to the mean sunspot number (SSN) binned over CR. There were 387 and 318 HCMEs during the study period in cycle 23 (1996 May 1 to 2006 Aug 31) and cycle 24 (2008 Dec 1 to 2019 Mar 31), respectively.

Table 1 shows the HCME properties in cycles 23 and 24 and in the corresponding phases of the two cycles. 'Cycle Duration' denotes the study period in the respective solar cycles. The starting dates in each cycle represent the beginning of the respective solar cycles which is a solar minimum. The end dates were set based on the latest recording of cycle 24 HCMEs by the catalog (https://cdaw.gsfc.nasa.gov/CME_list/halo/halo.html) and the corresponding epoch for cycle 23. 'Remainining in cycle 23' denotes the HCMEs outside the study period of cycle 23. The number of months without data refers to the period when SOHO was temporarily disabled (June to August 1998 and January 1999). The rate per month is the monthly rate of HCME occurrence. The mean SSN is the average value of the monthly mean total SSNs (http://www.sidc.be/silso/datafiles#total). The Rate/SSN is the ratio of the HCME rate per month to the mean SSN. 'Phase duration' represents the time period for the corresponding phases of the solar cycles. The rising phase starts from the solar mimimum (beginning of the solar cycle) and ends before the maximum phase starts. The monthly mean SSN smoothed by a 13 month running mean reached its minimum in May 1996 (cycle 23) and December 2008 (cycle 24). The maximum phase for cycle 23 started in the begning of 1999 which is evidenced by the arrival of high latitude ($\geq 60^0$) polar crown filaments and their rapid increase in the middle of 1999. The maximum phase ended in May 2002 which is indicated by the completion of polarity reversal at the solar poles (in the more active northern hemisphere and then months later in the sothern hemisphere). Then solar cycle 23 enters in to a long declining phase from the beginning of June 2002. The maximum phase for cycle 24 started in Feb 2011. There was a brief peak of activity in the northen hemisphere in 2011 and onother broader peak in the southern hemisphere in 2014. The first peak reached a maximum value of smoothed SSN in Feb 2012 and the second peak in April 2014 (the occurrence of the polar field reversal in the northern hemisphere). The late maximum occurs five years after the preceding minimum in December 2008. Then cycle 24 enters in to a long declining phase starting from May 2014.

**Table1:** Halo CME Properties in cycles 23 and 24 and in the corresponding phases.

| Halo CME Properties | | Solar Cycle | | |
|---|---|---|---|---|
| | | **Cycle 23** | **Cycle 24** | **Remaining in Cycle 23** |
| **Cycle Duration** | | 1996 May 1-2006 Aug 31 | 2008 Dec 1-2019 Mar 31 | 2006 Sep 1-2008 Nov30 |
| **Number of months** | | 124 | 124 | 27 |
| **Number of months without data** | | 4 | ---- | ---- |
| **Number of HCMEs** | | 387 | 318 | 9 |
| **HCME rate per month** | | 3.23 | 2.56 | 0.33 |
| **Mean SSN** | | 96.35 | 52.46 | 11 |
| **HCME rate/SSN** | | 0.034 | 0.049 | 0.03 |
| **Rise** | Phase duration | 1996 May 1-1998 Dec 31 | 2008 Dec 12-2011 Jan 1 | ---- |
| | Number of months | 32 | 26 | ---- |
| | Number of months without data | 3 | ---- | ---- |
| | Number of HCMEs | 49 | 12 | ---- |
| | Rate per month | 1.69 | 0.46 | ---- |
| | Mean SSN | 46.92 | 14.77 | ---- |
| | Rate/SSN | 0.036 | 0.031 | ---- |
| **Maximum** | Phase duration | 1999 Jan 1-2002 May 31 | 2011 Feb 1-2014 Apr 30 | ---- |
| | Number of months | 41 | 39 | ---- |
| | Number of months without data | 1 | ---- | ---- |
| | Number of HCMEs | 171 | 214 | ---- |
| | Rate per month | 4.28 | 5.48 | ---- |
| | Mean SSN | 161.93 | 91.88 | ---- |
| | Rate/SSN | 0.026 | 0.059 | ---- |
| **Decline** | Phase duration | 2002 Jun 1-2006 Aug 31 | 2014 May 1-2019 Mar 31 | ---- |
| | Number of months | 51 | 59 | ---- |
| | Number of months without data | ---- | ---- | ---- |
| | Number of HCMEs | 167 | 92 | ---- |
| | Rate per month | 3.27 | 1.56 | ---- |
| | Mean SSN | 74.65 | 43.02 | ---- |
| | Rate/SSN | 0.044 | 0.036 | ---- |

3.1.1. Intercycle comparison of HCME rates

The average SSN in cycle 24 declined by ~46%: 96.35 in cycle 23 versus 52.46 in cycle 24 averaged over 124 months. The HCME rate per SSN is higher in cycle 24: 0.049 in cycle 24 versus 0.034 in cycle 23. In spite of the significant SSN decline in cycle 24, the abundance of halo CMEs did not decline in cycle 24; the rate per SSN is in fact ~44% higher.

We compare the HCME rates in the corresponding phases (rise, maximum, and declining) of the two cycles as presented in Table 1. The absolute number of cycle-24 HCMEs in rise and declining phases are smaller than the corresponding numbers in cycle 23. On the other hand, the absolute number of HCMEs is higher in the maximum phase. The character of halo CMEs over the whole cycle is determined by the events in the maximum phase.

When monthly rates normalized to SSN are compared, the cycle-24 rates dropped by only 14% and 18% in the rise and declining phases. However, the SSN dropped by 68% and 42% in cycle 24. In the maximum phase, cycle-24 rate normalized to SSN increased by 127% while the SSN dropped by 43%.

The slightly higher HCME rate per SSN in the declining phase of cycle 23 is due to an unusual bump in HCME activity caused by abnormally energetic events occurring during October 2003 to October 2005. A detailed explanation is provided in section 4. In any case, the drop in cycle-24 monthly rate normalized to SSN is much smaller than the drop in SSN (42%) compared to the rate in the declining phase of cycle 23.

3.1.2. Intracycle comparison of HCME rates

We compare the HCME rates within the phases of each solar cycle. In cycle 24, the HCME rates per month and per SSN in the maximum phase are significantly higher than in the rise and declining phases. In cycle 23, the HCME rate per month in the maximum phase is significantly higher than in the rise and declining phases (see Table 1). However, the rate per SSN in the declining phase is higher than the rise and maximum phases. The unusuall bump in HCME activity (see section 4) accounts for this higher rate per SSN.

**3.2. Halo CME Source Locations**

Figure 2 shows the HCME source locations, distributions of longitudes and latitudes, and the variation of the source latitudes over time with the corresponding phases compared between the cycles. We use frontside halos because the source locations of cycle-23 backside halos are unknown. There were 242 frontside HCMEs (62.53% of the 387 HCMEs) in cycle 23 and 155 (48.74% of the 318 HCMEs) in cycle 24. The number of frontside HCMEs in the rise, maximum, and decline phases are respectively: 31, 114, and 97 in cycle 23; 10, 105, and 40 in cycle 24.

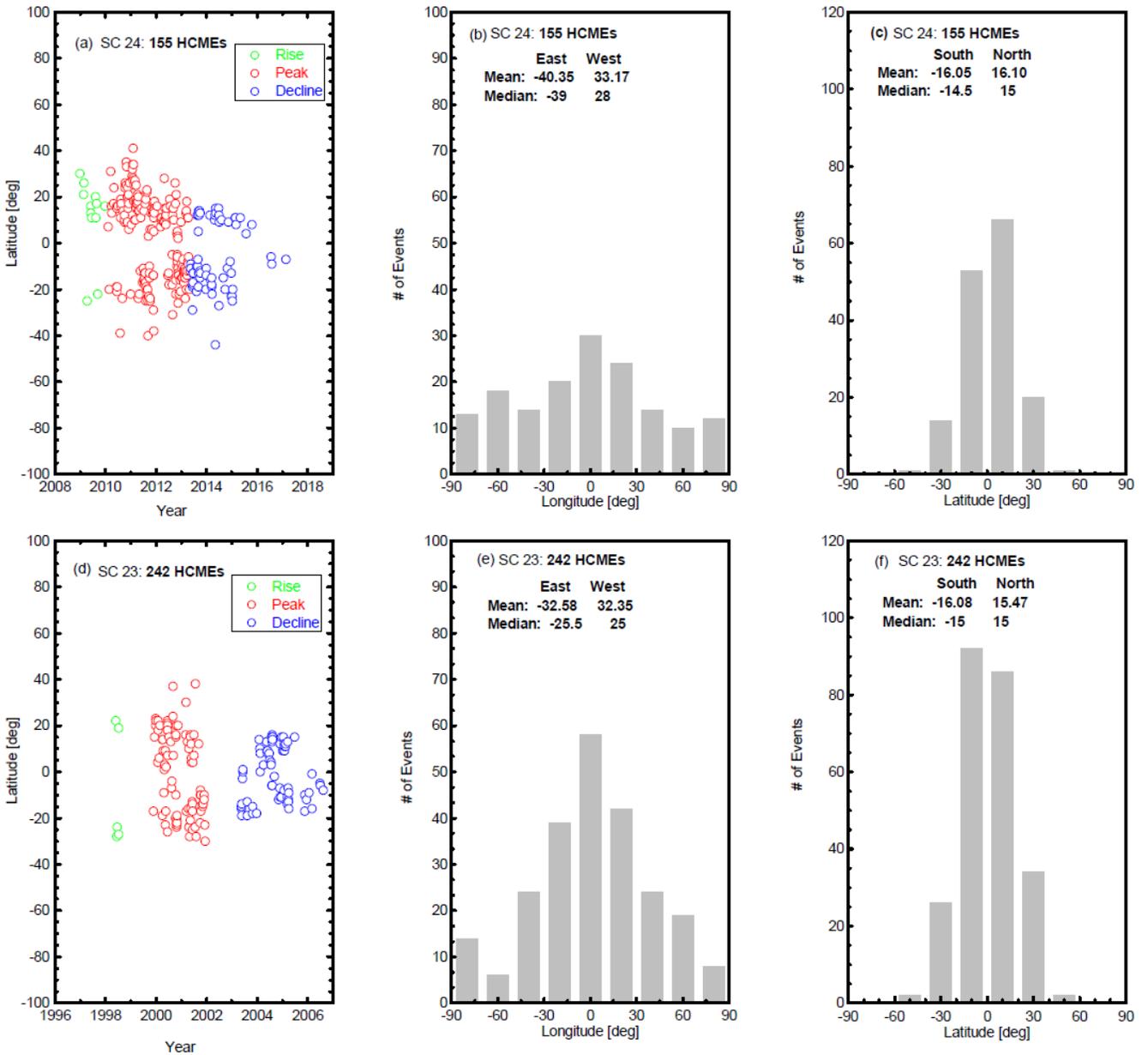

**Figure 2:** The solar source locations of HCMEs in cycle 24 (top row) and cycle 23 (bottom row). The left-hand panels (a) and (d) show the variation of the source latitude over time. The green, red, and blue circles refer to the rising, maximum, and declining phases in both cycles. The middle panels (b) and (e) show the distributions of the source longitudes and the right-hand panels (c) and (f) show the distributions of source latitudes. The mean and median values of the source longitude and latitudes are plotted on the panels. Each panel also has the total number of frontside HCMEs.

### 3.2.1. Solar-cycle variation of the source latitudes and longitudes

We found that 24% of the frontside HCMEs in cycle 24 and 18% of those in cycle 23 originate at a CMD $\geq 60^0$. About 58% of the front side HCMEs in cycle 23 and 49% of those in cycle 24 originate within a CMD of $30^0$.

Figures 2(a) and (d) show the solar cycle variation of the source latitudes over time. The distribution of HCMEs in latitude has peaks in the northern and southern hemispheres in both cycles which correspond to the active region belt. The high energy necessary for the energetic HCMEs is available in active regions. The HCME sources are progressively closer to the equator as the cycle progresses, resembling the butterfly diagram.

Figures 2(c) and (f) show the distributions of the source latitudes. There is no significant difference in the source latitude distributions between the two cycles. There are two distinguished peaks at $\pm 15^0$ in both cycles. We found that 119 Out of the 242 HCMEs (49.2%) in cycle 23 and 70 out of the 155 HCMEs in cycle 24 (45.5%) originate from $\pm(15^0$-$30^0$ latitudes).

The distribution of source longitudes in cycle 24 (Figure 2(b)) differ from cycle 23 (Figure 2(e)). The source longitudes of cycle 24 HCMEs are more uniformly distributed than those in cycle 23.

### 3.2.2. Phase-to-phase comparison of the source locations

There are very few frontside HCMEs in the rising phase of the two cycles, but cycle-24 has more events (10 out of the 12 HCMEs are frontside). This is opposite to what is listed for all HCMEs in Table 1. About 68% of the 155 frontside HCMEs in cycle 24 are at the maximum phase which is higher compared to 47% of the 242 frontside HCMEs in cycle 23. It appears that the set of HCMEs at the end of the declining phase of cycle 23 (Figure 2(d)) has no corresponding matches in cycle 24 (Figure 2(a)). This happens because of the exceptional active regions in different periods of the years 2003 to 2005 that produced an unusually large number of HCMEs (section 4 for details).

Table 2 shows a summary of the Kolmogorov-Smirnov (KS) test results for the HCME source locations and speeds. The p-value (p) refers to the probability that the difference between the distributions of the datasets is caused by random variations and not a statistical one. The smaller the p value,the stronger the evidence that the two data sets were sampled from populations with different distributions (we roughly consider $p \leq .05$ as 'significant', $p \leq .01$ as 'highly significant'.and $p > .10$ 'not significant'). D refers to the maximum absolute difference between the observed cumulative distribution functions of the two datasets. The critical value (D critical) is obtained using the formula $D_C = c(\alpha)\sqrt{\frac{n+m}{n.m}}$ where n and m are the sizes of the first and second data sets respectively. For a significance level α = 0.05, the value of c(α) = 1.36. If D > $D_C$, then it implies a significant difference between the distributions of the two datasets. There is no difference in the values of p, D, and $D_C$ between the sky-plane and space speeds in the rise, maximum, and declining phases.

**Table2:** Kolmogorov-Smirnov (KS) test result summary.

| HCME properties | Phase | p-value | Maximum difference between cumulative distributions(D) | Dcritical ($D_C$) |
|---|---|---|---|---|
| Source longitude | Rise | 0.99 | 0.14 | 0.48 |
| | Maximum | 0.027 | 0.19 | 0.18 |
| | Decline | 0.33 | 0.17 | 0.25 |
| Source latitude | Rise | 0.06 | 0.44 | 0.48 |
| | Maximum | 0.15 | 0.18 | 0.15 |
| | Decline | 0.06 | 0.24 | 0.25 |
| HCME speed | Whole cycle | 0.001(Sky-plane) | 0.15 (Sky-plane) | 0.1 |
| | | 0.003 (Space) | 0.14 (Space) | |
| | Rise | 0.77 | 0.2 | 0.44 |
| | Maximum | 0.34 | 0.095 | 0.14 |
| | Decline | 0.001 | 0.25 | 0.18 |

We performed a KS test for the phase-to-phase comparison of the source longitude distributions. The KS test in the maximum phase (D=0.19 > $D_{critical}$ =0.18 and p =0.027) implies significant difference between the two cycles (see Table 2). The rising (D=0.14 < $D_{critical}$ = 0.48 and p=0.99) and declining phases (D=0.17 < $D_{critical}$ = 0.25 and p=0.33) do not show significant difference between the two cycles. Therefore, we conclude that the different behavior of HCMEs in the two cycles is essentially determined by the HCMEs in the maximum phases. This result is consistent with the HCME rates discussed above (section 3.1.2).

We also conducted the KS test for the phase-to-phase comparison of source latitude distributions between the two cycles. The KS tests in the rising (D=0.44 < $D_{critical}$ = 0.48 and p=0.06), maximum (D=0.15 < $D_{critical}$ = 0.18 and p=0.15), and declining phases (D=0.24 < $D_{critical}$ =0.25 and p=0.06) show that the source latitude distributions are not significantly different.

### 3.3. Halo CME Speeds and Flare Size Distributions

In this section, we present a comparison of the sky-plane speed and space speed between the two cycles. In addition, we use the frontside halos and the associated flares to compare the flare size distributions.

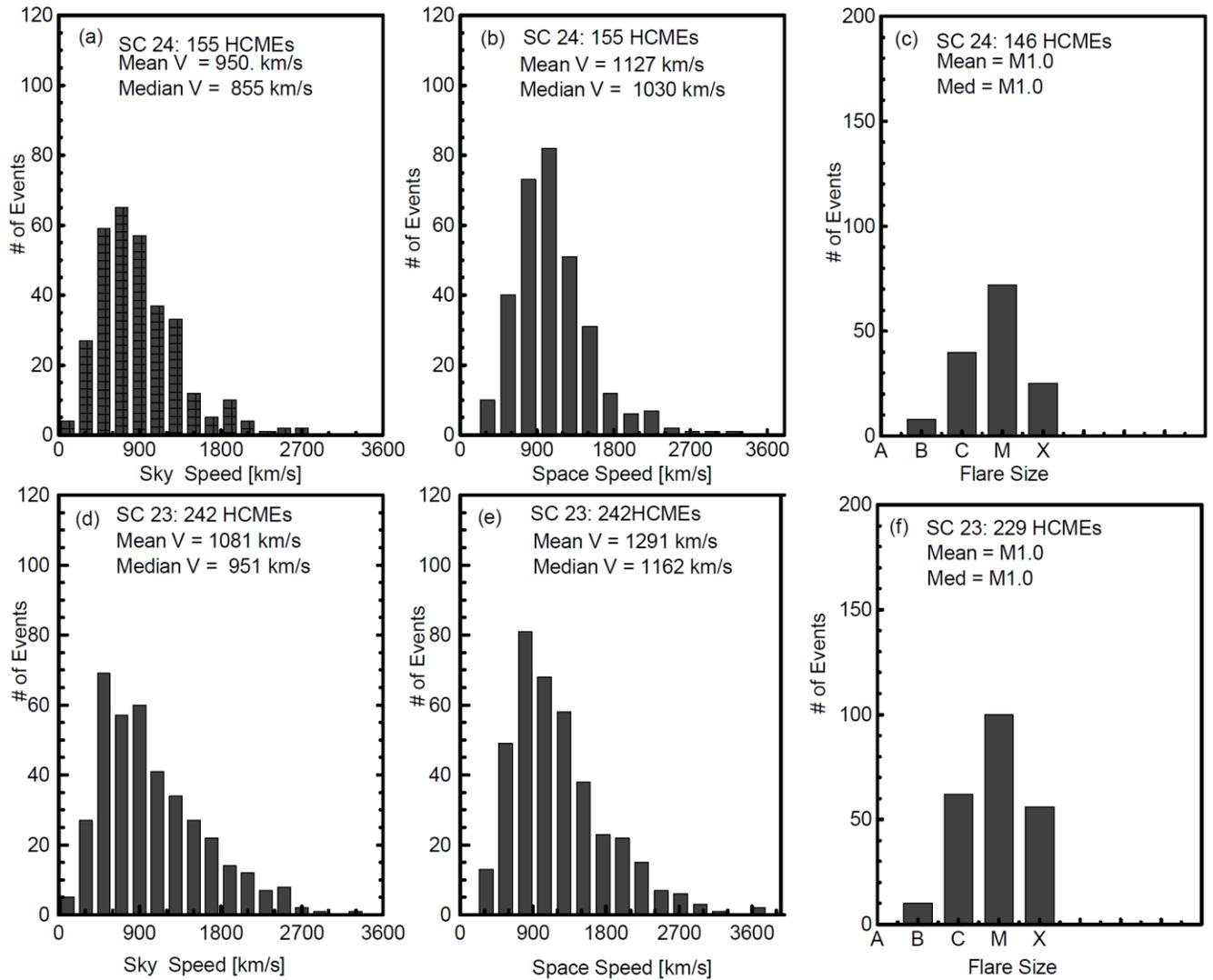

**Figure 3:** Comparison of speed and flare size distributions of frontside HCMEs in cycle 24 (top row) and 23 (bottom row). The distributions of the sky-plane and space speeds are shown in the left-hand and middle panels, respectively. The space speeds were calculated using the cone model. The right-hand panels show the size distributions of the associated flares. The number of HCMEs and the corresponding mean and median values are stated in each plot.

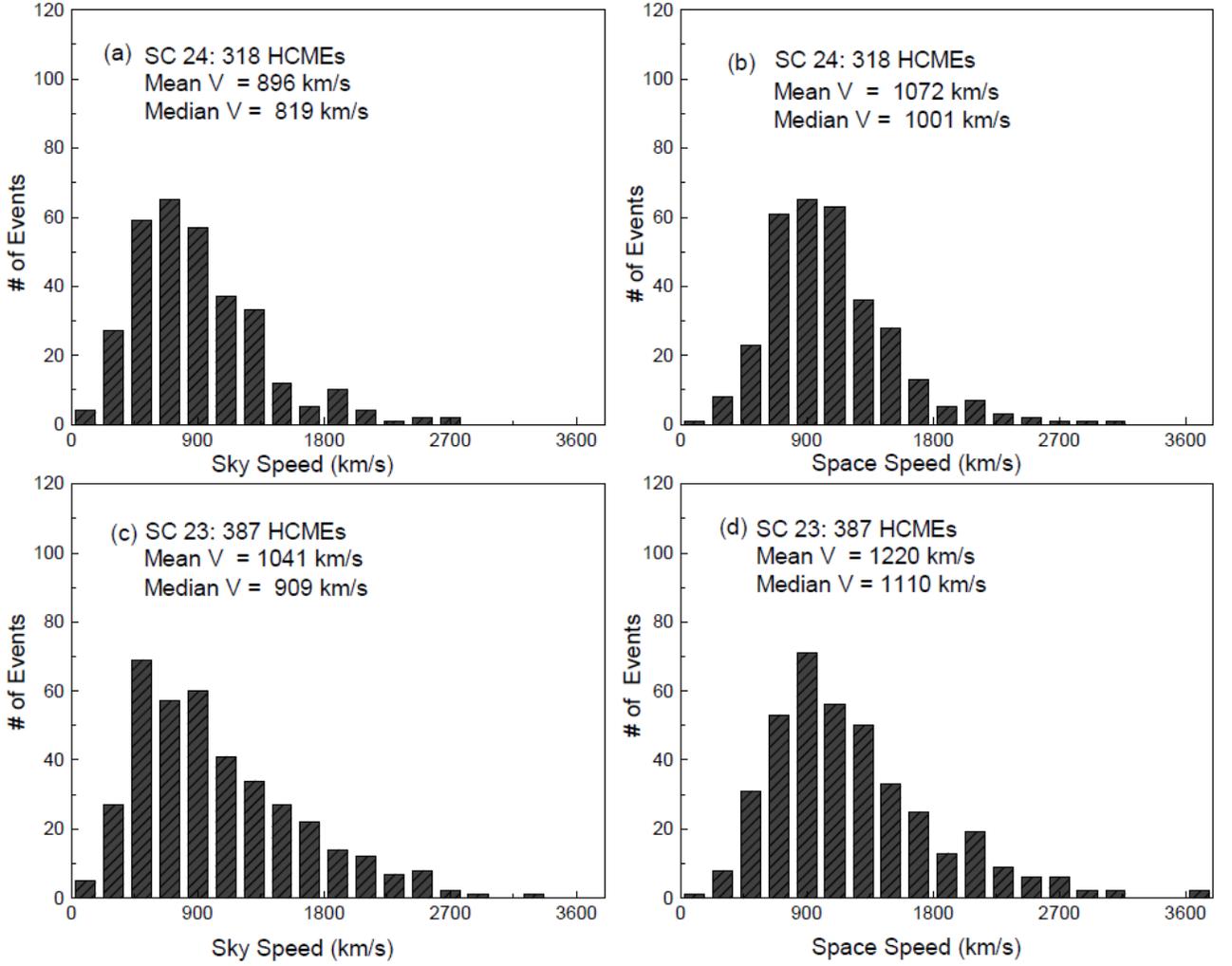

**Figure 4:** Comparison of the speeds of all HCMEs in the two whole cycles: cycle 24 (top row) and 23 (bottom row). The distributions of the sky-plane and space speeds are shown in the left-hand and right-hand panels, respectively. The mean and median values and the total number of HCMEs are plotted on the panels.

3.3.1. Kinematics and flare size distributions: comparison between the two cycles

The difference in the average speeds (sky-plane and space speeds) of the front side HCMEs between the two cycles is ~14% (See Figure 3). The mean values of the sky-plane and space speeds of the frontside HCMEs are: cycle 23 ($V_{sky}$ = 1081 km s$^{-1}$, $V_{space}$ = 1291 km s$^{-1}$) and cycle 24 ($V_{sky}$ = 950 km s$^{-1}$, $V_{space}$ = 1127 km s$^{-1}$). The difference in the average speeds increases when we consider all HCMEs between the two whole cycles. The mean values of the sky-plane and space speeds for the whole cycles are: cycle 23 ($V_{sky}$ = 1041 km s$^{-1}$, $V_{space}$ = 1220 km s$^{-1}$) and cycle 24 ($V_{sky}$ = 896 km s$^{-1}$, $V_{space}$ = 1072 km s$^{-1}$). The average sky-plane speed in cycle 23 is ~16% higher than in cycle 24 (See Figure 4).

The KS test for sky-plane speeds ( p = 0.001 and D = 0.15 > $D_{Critical}$ = 0.1) and space speeds ( p = 0.003 and D = 0.14 > $D_{Critical}$ = 0.1) implies that the speeds differ significantly. Gopalswamy et al.(2015a) did not find any speed difference when partial cycles were compared.

The flare size distributions are similar in the two cycles. The mean and median values of the soft X-ray maximum flux are the same (= M1.0) in both cycles. This is an indication that CME kinematics is affected by the heliosphere, while the flare sizes are not.

3.3.2. Sky-plane and space speed variations: comparision between the corresponding phases

We performed a KS test for the speeds of all HCMEs in the rising, maximum, and declining phases in the two whole cycles (See Figure 4 and Table 2). The KS tests for sky-plane and space speeds in the rise (D = 0.2 < $D_{critical}$ = 0.44, p= 0.77) and maximum (D = 0.095 < $D_{critical}$ = 0.14, p= 0.34) phases do not imply significant difference. In the declining phase (D = 0.25 > $D_{critical}$ = 0.18, p= 0.001); the test signifies that the speeds between the two cycles differ significantly.

Figure 5 shows the change in sky-plane speed with the CMD of the solar source for frontside HCMEs in cycles 23 and 24. The sky-plane speed generally increases as a function of the CMD while the space speed is independent of the CMD. The regression lines and the correlation coefficients are similar in both cycles. A significant difference between the sky-plane and space speeds is observed at a CMD < 45$^0$ whereas the speeds merge after CMD = 60$^0$. The average speed at CMD ≥ 60$^0$ for cycle 23 ($V_{sky}$ =1579 km s$^{-1}$, $V_{space}$ = 1596 km s$^{-1}$) is higher than that of cycle 24 ($V_{sky}$ =1238 km s$^{-1}$, $V_{space}$ = 1254 km s$^{-1}$). The average speed in cycle 24 is ~28 % smaller than that of cycle 23. Gopalswamy et al. (2020) recently reported that the average speed of limb HCMEs in cycle 24 is ~28% smaller than that in cycle 23. They suggested that cycle 24 CMEs become halos at a shorter distance from the Sun and at lower speeds whereas cycle 23 CMEs take too long and must be of higher speed to become halos. What we have found is in agreement with this report.

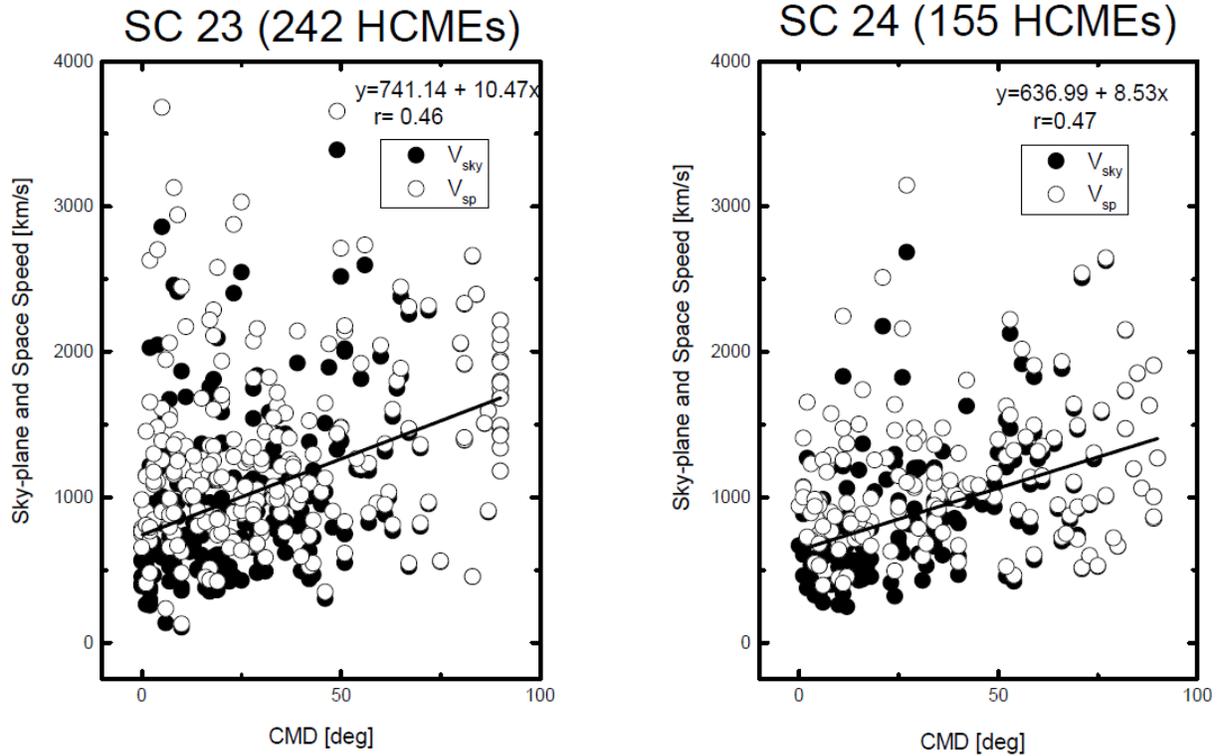

**Figure 5:** Scatterplot between HCME speeds and the CMD of the solar source in cycles 23 (left) and 24 (right). The sky-plane and space speeds for the corresponding cycles are represented by filled and open circles, respectively. The regression line and the corresponding equation together with the correlation coefficient (CMD vs. sky-plane speed) are shown in the plots. The difference between the sky-plane and space speeds of each cycle is large at smaller CMD but the speeds merge after CMD = $60^0$.

## 4. The unusual bump in HCME activity at the declining phase of solar cycle 23

An unusually large fraction of fast and wide CMEs and HCMEs occurred in cycle 23 which is reflected in the phase-to-phase comparison with cycle 24. There are 167 HCMES in the declining phase of cycle 23 (2002 Jun 1- 2006 Aug 31). The period from October 2003 to October 2005 accounts for 68% of the general population of HCMEs in the declining phase. This period constitutes nearly 28% of the frontside HCMEs in the entire cycle 23. A large number of fast and wide CMEs and HCMEs also occurred during this period. The mean and median values of the sky plane and space speeds in this period are higher than in the entire period of the declining phase. We found that in cycle 23, 35% of the front side HCMEs whose CMD $\geq 60^0$ belong to this specific period of the declining phase of cycle 23 which manifests a large fraction of fast HCMEs. About 40% of these HCMEs are with speeds $> 2000$ km s$^{-1}$.

We noticed an exceptionally large number of HCMEs (the HCME numbers are indicated in parenthesis alongside each period) in the following periods: 23 October to 5 November 2003 (13), 12 October to 8 November 2004 (9), 18 August to 14 September 2004 (7), January 2005 (9), 21 April to 18 May 2005 (8 ),14 June to 10 July 2005 (6), 11 July to 8 August 2005 (14), August to 4 September 2005 (8), and 05 September 2005 to 01 October 2005 (5). The HCME rate starts increasing in the year 2004 and peaks in 2005.

Gopalswamy et al. (2005) studied the CMEs and other extreme characterstics of the 2003 October to November solar eruptions. They reported an episode of intense solar activity characterized by the abundant occurrence of fast CMEs, X class flares, SEPs and interplanetary shocks. They found that 20% of the ultra fast CMEs and 12% of the HCMEs of the general population of cycle 23 occur during this period. They also stated that the rate of HCMEs for the specified period were nearly four times the average rate during cycle 23. Gopalswamy et al. (2006) studied the extreme solar activity and its space weather implications during the declining phase of the solar cycle 23 (October-November 2003, November 2004, January 2005, and September 2005). They reported that all eruptions are from superactive regions that are geoeffective and SEP productive. Although there are some differences between the four regions, they are essentialy characterized by large active region area, highest concentration of ultrafast CMEs and HCMEs.

The main reason for such extreme behavior is attributable to the exceptional super-active regions(SARs) with the largest extractable energy that produced an unusually large fraction of fast and wide CMEs and HCMEs.

In the same perspective,we also have to consider the extreme behaviors in the declining phase of solar cycle 24. We examine the active regions associated with extreme solar activities (the release of X class flares, SEPs and GLE events) in the declining phase for CME productivity. The number of HCMEs associated with the specified periods of the active regions are: AR NOAA 12673 (September 4-12, 2017: 3 HCMES), AR NOAA 12371 ( June 18-25, 2015:4 HCMEs), and AR NOAA 2297 (March 11-17,2015:1 HCME). Chen and Wang (2016) studied the characterstics of SARs in solar cycle 24 until August 31,2015. They reported that the velocities of CMEs associated with SARs in cycle 24 are lower compared to the preceeding cycles. The SARs, although flare rich, they are CME poor. For example, AR NOAA 12192 ( October 17-30, 2014) didn't produce any significant earth directed CME. Among tens of major flares, only an M4.0 flare was associated with a CME. Therefore, there are no significant numbers and speeds of HCMEs due to extreme behaviors in the declining phase of solar cycle 24.

## 5. SUMMARY AND CONCLUSIONS

We have made comparisons of the HCME properties between cycle 23 (1996 May 1 to 2006 Aug 31) and cycle 24 (2008 Dec 1 to 2019 Mar 31) during the first 124 months in each cycle. The importance of this work in light of the previous study (Gopalswamy et al.(2015a) can be explained as: I. The declining phases of the solar cycles which were not part of the previous study are included. Besides, solar cycle 24 has come to its end that enables us to perform whole cycle comparsions. II. In addition to comparing the whole cycles, the new work compares the HCME properties in the corresponding phases (rise, maximum, and declining) in the two cycles. III. Automatic detection catalogs have recorded too few halo CMEs, this whole-cycle HCME comparison will serve as a bench mark to compare HCMEs from automatic detection algorithms. By considering a much larger data sets in two whole solar cycles, we have confirmed the findings by Gopalswamy et al.(2015a). The specific findings of our study can be stated as follows:

1. In comparing the whole cycles, we see that there were 387 HCMEs in cycle 23 and 318 in cycle 24. The SSN averaged over 124 months dropped by 46% in cycle 24 with respect to cycle 23. Despite the significant decline in SSN in cycle 24, the abundance of HCMEs is similar in the two cycles. The rate per SSN in cycle 24 is infact ~44% higher.
2. Although the SSN in the maximum phase of cycle 24 drops off by 43%, the HCME rate per SSN increased by 127%. The cycle-24 rates per SSN dropped by only 14% and 18% in the rise and declining phases while the SSN dropped by 68% and 42%, respectively.
3. A comparsion of the HCME rates within the phases of each solar cycle shows that the HCME rates per month in the maximum phase are significantly higher than in the rise and declining phases.

4. The source longitudes of cycle 24 HCMEs are more uniformly distributed than those in cycle 23. The source longitude distributions do not differ significantly in the rise and decline phases between the two cycles. However, at the maximum phase, they are significantly different.
5. The distribution of HCMEs in latitude has peaks in the northern and southern hemispheres in both cycles which correspond to the active region belt. The high energy necessary for the energetic HCMEs is available in active regions. The HCME sources are progressively closer to the equator as the cycle progresses, resembling the butterfly diagram. There is no significant difference in the source latitude distributions between the two cycles and the corresponding phases.
6. The different behavior of HCMEs in the two whole cycles is essentially determined by the character of HCMEs in the maximum phases.
7. We found that the percentage of HCMEs originating at a CMD $\geq 60^0$ is higher in cycle 24 than in cycle 23. A significant difference between the sky-plane and space speeds of each cycle is observed at a CMD $< 45^0$ whereas the speeds merge after CMD $= 60^0$.
8. The average speed (sky-plane and space speed) at CMD $\geq 60^0$ for cycle 23 is ~28% higher than that of cycle 24.
9. The average sky-plane speed in cycle 23 is ~16% higher than in cycle 24. The HCME kinematics between the two cycles differ significantly. The size distributions of the associated flares between the two cycles and the corresponding phases are similar. This is an indication that CME kinematics is affected by the heliosphere, while the flare sizes are not.
10. The peculiarities in the phase-to-phase comparison are affected by the extreme behaviors (large fraction of HCMEs, high HCME rates, and very fast CMEs) manifested at the declining phase of cycle 23. This happened due to an unusuall bump in HCME activity caused by exceptional active regions that produced a large number of CMEs during October 2003 to October 2005.

The total pressure, magnetic field, and density significantly diminish in the inner heliosphere due to low solar activity in cycle 24 as compared to cycle 23. Hence, the state of the heliosphere is significantly weaker in cycle 24. We conclude that the anomalous expansion of cycle 24 CMEs accounts for the general increase in HCME abundance and the increased number of halos at larger CMD. Furthermore, the maximum phase shows the highest contrast between the two cycles.

## ACKNOWLEDGMENT

We acknowledge NASA's open data policy in using SOHO and STEREO. SOHO is a project of international collaboration between ESA and NASA. STEREO is a mission in NASA's Solar Terrestrial Probes program. This work is supported by NASA's Living With a Star program. FD thanks the Ethiopian Space Science and Technology Institute and the Mekelle University for partial financial support.